\documentstyle[psfig]{caosp}
\begin{document}
\def\teff{$T_{\rm eff}$}
\def\lgg{$\log\,{g}$}
\def\vt{$\xi_{\rm t}$}
\def\vsini{$v\cdot \sin i$}
\def\kms{\,km\,s$^{-1}$}
\def\gequ{$\gamma$\,Equ}
\def\alcir{$\alpha$\,Cir}
\def\i{\,{\sc i}} \def\ii{\,{\sc ii}} \def\iii{\,{\sc iii}}
\def\kG{\,kG}
\def\vsini{$v\cdot \sin i$} \def\hfs{hyperfine-structure\ }


\htitle{CP star atmospheres based on individual ODFs}
\hauthor{F. Kupka and N.E. Piskunov}
\title{CP star atmospheres based on individual ODFs}
\author{F. Kupka \inst{1}  \and N.E. Piskunov \inst{2}}
\institute{Institute for Astronomy, University of Vienna, Vienna, Austria
\and Uppsala Astronomical Observatory, Uppsala, Sweden}
%
\maketitle

\begin{abstract}
We describe a new method for the computation of opacity distribution 
functions (ODFs) useful to calculate one-dimensional model atmospheres in
local thermal equilibrium (LTE). The new method is fast enough to be
applied on current workstations and allows the computation of model
atmospheres which deviate significantly from (scaled) solar chemical
composition. It has reproduced existing ODFs and model atmospheres for
solar abundances. Depending on the type of chemical peculiarity the
``individual'' model atmosphere may have a structure and surface fluxes
similar to atmospheres based on (scaled) solar abundances or
deviate in a way that cannot be reproduced by any of the conventional
models. Examples are given to illustrate this behavior. The availability
of models with ``individualized'' abundances is crucial for abundance
analyses and Doppler imaging of extreme CP stars.
\end{abstract}

\section{Introduction}

Opacity distribution functions are used to describe line opacities
$l_{\nu}$ in LTE for a set of pairs of temperature T and gas pressure P
assuming a fixed chemical composition and (in our case) a fixed value
for the microturbulence $\xi_t$:
\begin{equation}   \label{kupka_eq1}
  l_{\nu} = l_{\nu}(T,P,{\rm chem.~composition},\xi_t).
\end{equation}
The frequency dependence of the $l_{\nu}$ is described by dividing the
whole wavelength range relevant to the desired stellar atmosphere
computations into 300 to 1200 channels. Each of these channels is devided
into 10-12 subchannels which provide a statistical representation of line
opacity (ranging from high to low values). The (radiative) flux integrated
over a set of subchannels approximates the overall (radiative) flux for
each particular channel. Thus, one can compute the total radiative flux
throughout the model atmosphere, as well as surface fluxes and 
intensities. This technique was described in Strom
and Kurucz (1966) and later used, for instance, by Gustafsson et al. (1975)
and by Kurucz (1979). 

\section{Description of the new method} 

Muthsam (1979) was the first to compute model atmospheres for CP
stars. In the mean time, the reliability of atomic data obtained
from experiments has been improved. Similarly, available line lists have
increased both in size and reliability by up to an order of magnitude.
Hence, it is worthwhile to bring individual model atmospheres for CP stars
to the standards for stars with solar elemental abundance (cf.\ 
Kurucz 1993).

For his work Muthsam (1979) used the opacity sampling (OS) technique
which requires the computation of a crude synthetic spectrum. The latter
has to be sufficiently accurate for both the calculation of the integral 
flux in a wavelength region similar in size to the channels used for the 
ODF technique as well as for the calculation of the total radiative flux 
in each model layer. The OS technique is more efficient for the computation 
of {\em single} models. As opposed to the ODF technique, opacity sampling 
allows the study of vertical stratification of elemental abundances.
Moreover, the OS technique is also capable of representing the blanketing
effect in cool stars where the wavelength distribution of the opacities
may change dramatically with depth as new molecules are formed (cf.\
Ekberg et al. 1986). However, as we did not intend to study stratification
or late type stars with our code, we rather decided to benefit from the main
advantage of the ODF technique: the rapid computation of small {\em grids}
of model atmospheres (as a function of \teff\ and \lgg). Such grids are
very convenient for spectroscopic analyses, the computation of color
and flux grids, the investigation of the flux distribution of pulsating
stars, and Doppler imaging.

Speed and accuracy requirements for an ODF computation are determined by
a) the number of T-P pairs in (\ref{kupka_eq1}) used to represent the 
whole ODF, b) the number of spectral lines used for the opacity
computation, and c) the size of the wavelength grid used for each 
channel ``$l_{\overline{\nu}}$''. Under the assumption that the structure
of the model atmosphere does not change dramatically for a different
chemical composition, we can adapt our computations according to each of
these three quantities. First, the T-P pairs are selected from a model
atmosphere that is closest in \teff\ and \lgg\ as well as chemical
composition to the ``target'' model atmosphere (or model grid within some
limited \teff\ and \lgg\ range). In a second step, we use the Vienna
Atomic Line Data Base (VALD, Piskunov et al. 1995) together with its
extraction tools PRESELECT and SELECT to choose between 70\,000 and
600\,000 lines (a range valid from $\lambda$ Boo stars to extreme Si
peculiar stars) out of 42 million lines. The lines are selected according
to wavelength range, ionization stage, excitation potential, and the ratio
of line vs.\ continuous opacity for the desired range of T-P pairs or a
set of T-P pairs taken from the ``closest'' standard model atmosphere.
Finally, the ODF as well as Rosseland and/or Planck
mean opacities are computed in a two-stage process: an adaptive wavelength
grid (similar to that one for SYNTH as described in Piskunov 1992)
is used by the VOPDF code to compute line opacities for each T-P pair.
These are processed by the OPDF code which uses an adaptive histogram
and a running geometric mean to compute the desired opacity tables.
Details will be given in Piskunov \& Kupka (1998). Currently, the equation
of state and the continuous opacities are taken from ATLAS9 as published by
Kurucz (1993) to simplify the comparison with standard models. Without
line extraction, a typical ODF computation takes between 2 and 24 hours on
an Alpha workstation with a DEC-21164/A CPU running at 600 MHz (or between
6 and 72 hours for a DEC-21064/A at 266 MHz). The VOPDF/OPDF codes can
immediately run in parallel as opacities for different T-P pairs do not
depend on each other.

\section{Comparisons and Results}

\begin{table}[t]
\caption{Parameters used for ODF and model atmosphere computations.}
\label{Kupka_tab1}
\begin{center}
\begin{small}
\begin{tabular}{lrllr}
\noalign{\smallskip}
\hline\hline
  Star name  & \teff[K] & \lgg & $\xi_t$[km/s] & chem.~composition as in \\ 
\hline
    Vega     &    9550  & 3.95 &      2        & Castelli \& Kurucz (1994) \\
$\alpha$ Cir &    7900  & 4.2  &     1.5       &  Kupka et al. (1996) \\
  ET And     &   11500  & 3.6  &      2        &  Kuschnig et al. (1995) \\
\hline\hline
\end{tabular}
\end{small}
\end{center}
\end{table}

We present here some of the results of opacity calculations for
three different CP stars. For the first case, Vega, we compared available
{\em observations} of its flux distribution from the Lyman to the Paschen
series with surface fluxes derived from our own, {\em individual model
atmospheres} for Vega (see Table~\ref{Kupka_tab1}) as well as with fluxes
derived from a set of {\em standard model atmospheres} assuming a scaled
solar abundance (by -0.5 dex, using an ODF published on CDROM~3 of
Kurucz). The stellar parameters and chemical compositions were taken
from Castelli \& Kurucz (1994). The model atmosphere computations
were done with different variants of the ATLAS9 code of Kurucz. Details
will be given in Piskunov \& Kupka (1998). The overall agreement between
observations, individual models, and standard models is quite good. The
fluxes of individual and standard models are usually closer to each
other than to the observations. Small differences between
the ``standard models'' and ``individual models''
mainly originate from differences in the gf-values of the line lists used,
a different treatment of hydrogen lines, and from the fact that we used
the individual chemical composition determined for Vega. The latter
does not change the model structure, but generates some specific
flux features. This is supported by the result that the standard and
the individual model atmospheres deviate by less than 30 K for all
layers within the continuum as well as within the line formation region. 
Numerical experiments provide no evidence for a difference between
individual models created from ODFs with 1\% or 0.1\% as a minimum ratio
of line vs.\ continuum opacities. Hence, we used the larger minimum ratio
for most of the other ODF computations to save CPU time.

\begin{figure}
\centerline{
\psfig{figure=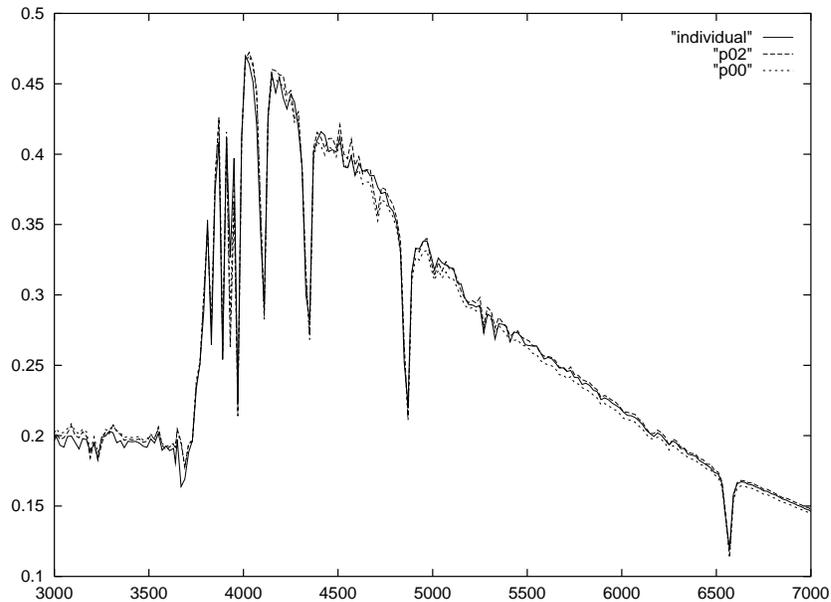,width=11.5cm}}
\caption{Flux distribution from model atmospheres for $\alpha$ Cir:
   individual (chemical composition as in Kupka et al. 1996), solar
   scaled by +0.2 dex, and solar. $F_{\lambda}$ is plotted here as a
   function of wavelength in {\AA} and scaled by a factor of $10^7$.} 
\label{Kupka_fig1}
\end{figure}

As a second case, we compared model atmospheres based on Kurucz'
ODFs for scaled solar abundances (by -0.2 dex, 0.0 dex, and +0.2 dex)
with models based on an individual ODF with an abundance pattern as
described in Kupka et al. (1996) for the mildly overabundant roAp star
$\alpha$~Cir (cf.\ Table~\ref{Kupka_tab1}). Here, the T vs.\ optical depth 
and the T vs.\ P relations fall essentially between the scaled
solar abundance patterns with 0.0 dex and +0.2 dex.
Hence, the line profiles remain unaffected by the new model
atmosphere. The behaviour of the flux distribution is more complicated and
is shown in Figure~\ref{Kupka_fig1}. Though Fe appears to be slightly
underabundant in $\alpha$ Cir, the individual model atmosphere of this star
produces a flux distribution closer to the case of an overabundance
of +0.2 dex, a slightly enhanced UV line blanketing after the Balmer jump,
various features of Sr~{\sc ii} (at 4077~{\AA} and at 4215~{\AA}) and a
mild depression at 5200~{\AA}.

Finally, we computed ODFs with different He, Si, and Fe abundances
representing the mean composition and abundance spots of the strongly Si
overabundant star ET And and compared the individual models with those
computed under the assumption of a scaled solar abundance (+0.0 dex
and +1.0 dex). Already for a mean overabundance of Si of +1.0 dex,
deviations from the ``standard models'' occur that cannot be ``simulated''
by changing \teff, \lgg, or by choosing a specific scaling factor for
the abundance. Though a specific feature or a single layer may be
``fitted'' easily, one cannot match the height of the Balmer discontinuity, 
or the relation of T vs.\ optical depth, or the flux distribution in the
visual and the UV at the same time for this star, and abundance analyses
based on scaled solar abundance patterns suffer from systematic errors.

\vspace{-2mm}
\section{Summary}

We have presented a new method for the computation of model atmospheres
for CP stars based on the ODF approach. The method was successfully
compared with standard models. Application to various CP stars shows that
model atmospheres for some abundance patterns ($\lambda$ Boo stars or
mildly peculiar roAp stars) can be computed using scaled solar abundances.
For other patterns, systematic deviations occur which cannot be
approximated by choosing a model atmosphere based on properly scaled solar
abundances, making the computation of individual ODFs (or model
atmospheres based on the OS technique) mandatory for applications such as
abundance analyses and Doppler imaging. One prominent example are Si
peculiar stars, essentially because Si is genuinely abundant and
many of its absorption lines cluster in certain wavelength regions. More
details on this work will be given in Piskunov \& Kupka (1998).
 
\acknowledgements
International cooperation and hardware for computations were supported by
the Fonds zur F\"orderung der Wissenschaftlichen Forschung (project
S7303-AST) who also provided funds for F. Kupka. We are grateful
to R. L. Kurucz for his permission to use ATLAS9 and line lists from
his CDROM distribution.

\end{document}